\documentclass[aps,pre,twocolumn,showpacs]{revtex4}
\usepackage{epsfig}
\usepackage{times}
\begin{document}

\title{Phase transitions for rock-scissors-paper game on different networks}

\author{Attila Szolnoki and Gy\"orgy Szab\'o}
\affiliation
{Research Institute for Technical Physics and Materials Science
P.O. Box 49, H-1525 Budapest, Hungary}

\begin{abstract}
Monte Carlo simulations and dynamical mean-field approximations are 
performed to study the phase transitions in rock-scissors-paper 
game on different host networks. These graphs are originated from lattices 
by introducing quenched and annealed randomness simultaneously.
In the resulting phase diagrams three different stationary states are 
identified for all structures. The comparison of results on different 
networks suggests that the value of clustering coefficient plays an
irrelevant role in the emergence of a global oscillating phase. The
critical behavior of phase transitions seems to be universal and can
be described by the same exponents. 
\end{abstract}


\maketitle

Complex networks are recently studied extensively as these objects
pervade all nature \cite{albert:rmp02,dorogovtsev:03,newman:siam03}. 
Initial efforts were focused on the characterization and evolution of
these graphs. Very recently the investigations are extended to critical 
behavior in dynamical models defined on various random networks 
\cite{medvedyeva:pre03,klemm:pre03}.
The effects of the network topology on critical transitions can well be
investigated on the small-world networks introduced by Watts and
Strogatz \cite{watts:nat98} because this structure provides a transition
from a regular lattice to some random networks. 
It is shown that the modifications in the host topology either change the
class of universality \cite{campos:pre03,graham:pre03}
or the long-range connections prevent complete ordering \cite{castellano:epl03}. 

Besides it, there are examples where the increase of randomness of the host
network induces a phase transition towards a state that cannot be observed
on lattices. To be specific, the emergence of an oscillating state has been
observed by Kuperman and Abramson \cite{kuperman:prl01} at a finite rate of
randomness in an epidemiological model. The appearance of global oscillation 
(synchronization) is also
observed in some cyclically dominated three-state systems, such as the
voluntary prisoner's dilemma \cite{szabo:pre04} and rock-scissors-papers (RSP)
games \cite{szabo:jpa04}. For the latter models the amplitude of oscillation
depends on the randomness and a second transition may occur when the limit
cycle approaches the absorbing states. The application of annealed (temporal)
randomness results in qualitatively similar behaviors. In this Brief Report,
concentrating on RSP game, we extend our previous work by combining both
types of randomness to explore the global phase diagram and to clarify
a universal feature.
 
For the spatial rock-scissors-paper game the individuals located on site
$i$ of a lattice belong to one of the three species ($s_i=1,2,3$) which dominate
cyclically each other. This means that the time evolution of this system is
governed by the iteration of cyclic invasion processes between two randomly
chosen neighboring sites. Namely, the pairs ($1,2$) or ($2,1$) become ($1,1$),
($2,3$) or ($3,2$) become ($2,2$), and finally ($3,1$) or ($1,3$) evolve
to ($3,3$) with the same rate defined to be 1. Starting from a random initial
state on a square lattice this system evolves into a self-organizing pattern
in which the three species are present with the same average concentration
($1/3$). This state maintained by the cyclic invasions 
\cite{tainaka:pla93,frachebourg:jpa98,ifti:epje03} provides protection against 
some types of invaders \cite{boerlijst:pd91,szabo:pre01b} and exhibits an 
unusual response to the external perturbation \cite{tainaka:epl91,frean:prs01}.

In our previous paper \cite{szabo:jpa04} the RSP game is studied on 
such regular small-world structures where randomly chosen (long-range)
links are substituted for portion $Q$ of the nearest-neighbor links
between sites located on a square lattice. A similar algorithm was proposed by
Watts and Strogatz \cite{watts:nat98} as a continuous transition between a
lattice and a random network. For the sake of simplicity, our rewiring process 
\cite{szabo:jpa04} conserves
the regularity, i.e. the coordination number of each site (called degree) remains
unchanged ($z=4$). Evidently, for $Q=0$ this structure is equivalent to
a square lattice meanwhile the limit $Q \to 1$ results in a regular random
graph. It is important to note that the restriction of fixed degree does
not change the small-world feature of the network: the characteristic (average)
path length between two sites scales as the logarithm of the network size.
Consequently the main conclusion is expected to be also valid for the standard
Watts-Strogatz networks. At the same time, this simplification has allowed us
to compare the results from these quenched structures with those obtained from
''annealed'' structures that can also be investigated analytically. For the 
annealed structures the (long-range) links are substituted temporarily with a
probability $P$ for the standard (nearest-neighbor) links on the square lattice.
Naturally, in the limit $P \to 1$ the dynamics of the system satisfies the 
mean-field condition. The effect of annealed randomness on rumor propagation 
\cite{zanette:pre01,zanette:pre02} and for the two-state voter model
\cite{vilone:pre04} has already been studied. 

For the RSP model on the quenched structures the emergence of global oscillation
is observed if $Q$ exceeds a threshold value and the amplitude of oscillation
tends to a fixed value in the limit $Q \to 1$. Such a transition occurs also
for the annealed structures when increasing the value of $P$. In the latter case,
however, the amplitude of global oscillation increases with time and finally
the evolution ends in one of the homogeneous (absorbing) states if $P$ exceeds 
a second threshold value. The different consequences of the two types of
randomness have inspired us to combine these randomness when starting from
different lattices. 
 
Henceforth, both types of randomness are applied to derive a global phase 
diagram on the $P-Q$ plane.
To explore the robust (universal) behaviors we also analyze
what happens for other regular structures with degree of $z=3$, 4, and 6.
These lattices are Kagom\'e (where $z=4$ as for the square lattice), 
honeycomb ($z=3$), triangle ($z=6$), cubic ($z=6$), and ladder-shape ($z=3$) 
lattices. For the ladder-shape structure two parallel chains are connected
by interchain bonds.

Our analysis is based on systematic Monte Carlo (MC) simulations.
First we create a quenched random regular structure starting from one of
the above mentioned lattices. That is, a portion $Q$ of
the nearest-neighbor links are replaced by randomly chosen links in a way
that  conserves the regularity (for details see \cite{szabo:jpa04}).
The MC simulations are started from a random (uncorrelated) initial state
where the three species take their place with the same probability (1/3).
Keeping this structure fixed the time evolution is governed by 
invasions between neighbors with a probability $1-P$ or along a (random) long
range link chosen with a probability $P$. In the simulation the number of
lattice points is varied from $10^3$ to $10^7$. The large sizes are used in
the close vicinity of transition points to reduce the undesired effect of
fluctuations.

\begin{figure}
\centerline{\epsfig{file=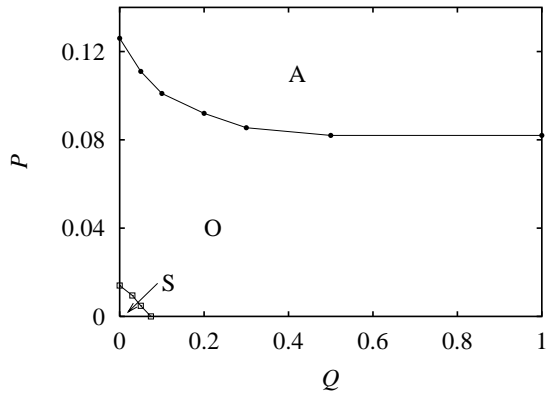,width=8cm}}
\caption{\label{fig:mfig1}Phase diagram for the RSP game on network
derivated from honeycomb lattice. The system exhibits self-organizing ($S$), 
oscillating ($O$), and absorbing (A) phases. Symbols represent MC results. 
The solid lines are guides to the eye.}
\end{figure}

The main feature of the steady-state phase diagram, which is generally valid
for all structures, can be summarized as follows. For small values of $Q$
and $P$, the stationary-state is characterized by a self-organizing strategy
distribution denoted by $S$. In this self-organizing pattern the strategies
alternate cyclically at each site, these local oscillations are not
synchronized by the short range interactions, and the average concentrations
are the same (1/3). For the opposite limit - when both structural parameters,
$Q$ and $P$, are close to $1$ - the system evolution is characterized by growing
spiral trajectories \cite{szabo:jpa04} and finally the evolution ends in one of
the three absorbing (homogeneous) states ($A$) containing only one strategy.
Evidently, this absorbing phase is three-fold degenerated due to the cyclic
symmetry. These two phases ($S$ and $A$) are separated by the region of global
oscillation ($O$) on the $P-Q$ plane. In this oscillating phase the behavior
is characterized by a limit cycle which is quantified by an order parameter
$\Phi$ defined as the ratio of the area of the limit cycle and area of the full
triangle in the ternary phase diagram \cite{szabo:jpa04}. This order parameter is
0 (1) in the state $S$ ($A$) and varies from 0 to 1 for the occurrence of
global oscillation (phase $O$). Figure \ref{fig:mfig1} shows the phase
boundaries obtained by MC simulations for $z=3$ on the $P-Q$ plane. 

\begin{figure}
\centerline{\epsfig{file=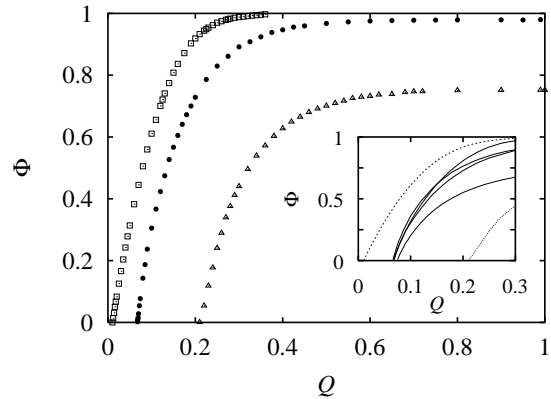,width=8cm}}
\caption{\label{fig:mfig2}The order parameter as a function of $Q$ for 
quenched networks ($P=0$) derived from cubic (box), square (bullet), and 
ladder-shape lattice (triangle). The corresponding critical value of 
$Q$ where the oscillatory phase emerges is $Q_1(P=0)=0.0103(1)$, $0.067(1)$,
and $0.210(2)$ respectively. The curves from left to right in the inset
show the $\Phi(Q,P=0)$ functions for small-world networks derived from
cubic, triangle, square, Kagom\'e, honeycomb, and ladder-shape lattices.}
\end{figure}

A striking quantitative difference is found in the behavior of these systems
when varying the value of $z$. Namely, global oscillation
(phase $O$) can be observed on random regular graphs ($Q=1$, $P=0$)
for $z=3$ and $z=4$. On the contrary, the system evolution terminates in
one of the absorbing states ($A$) on random regular graphs for $z=6$
and similar behavior is expected for $z>6$. In order to illustrate the
relevance of $z$ Fig.~\ref{fig:mfig2} shows the $Q$-dependence of
$\Phi$ for quenched structures ($P=0$) created by the rewiring technique
from the above mentioned lattices.  

Evidently, in the limit $Q \to 1$ the quenched structures become independent
of the original lattice. Consequently, starting from either the square or 
the Kagom\'e
lattices the order parameter $\Phi(Q,P=0)$ tends monotonously to the same
value ($\Phi_1(z=4)=0.980(5)$) in the limit $Q \to 1$. For $z=3$ (triangular
and ladder-shape lattices) we have obtained a lower limit value, namely
$\Phi_1(z=3)=0.750(5)$. For $z=6$, however, the system ends in the phase
$A$ if $Q$ exceeds a threshold value $Q_2$. According to our simulations
on the quenched structures ($P=0$) $Q_2=0.378(8)$ if the random regular
structure is created from a cubic lattice and  $Q_2=0.405(10)$ for the
triangular lattice. 

The topological structure of the original lattice affects the value of
$Q_1$ because the global oscillation occurs for small $Q$ for all the
investigated lattices. The inset in 
Fig.~\ref{fig:mfig2} demonstrates clearly that the lowest $Q_1$ appears 
for the three-dimensional cubic lattice, while the highest value is
found for the one-dimensional ladder-shape structures. 
It is underlined that the values of $Q_1$ are very close to each other 
for all the small-world structures created from the two-dimensional lattices.

The present RSP system undergoes two subsequent phase transitions when
increasing the randomness ($Q$ and/or $P$) of the backgrounds. The first
transition (from $S$ to $O$) is a Hopf-bifurcation that is well studied
by mean-field type approaches \cite{hofbauer:98}. Notice that the order
parameter $\Phi$ vanishes linearly for all the structures (see 
Fig.~\ref{fig:mfig2}) in agreement with the theory.

Our results indicate clearly that the global oscillation emerges just
above a threshold value of quenched randomness although the small-world
feature characterizes this type of networks at any small rate of disorder. 
The same phenomenon was observed by Kuperman and Abramson
\cite{kuperman:prl01} when considering the transition to an oscillating
phase in a three-state (susceptible-infected-refractory) epidemiological
model where the initial lattice was the traditional one-dimensional 
ring. They have conjectured that the emergence of global oscillation is
related to the variation in the clusterization. In the present model,
however, our results suggest that the clustering coefficient ($C$) cannot
play significant role because the corresponding values of $C$ are very
different for the studied structures. We cite as an example the case where
$C=0$ for all initial lattices except the triangle and Kagom\'e where
$C=2/5$ and $1/3$ respectively. Furthermore,
the value of C is ranging from $0.0004$ to $0.3$ for different
topologies when the system enters into the global oscillating phase.
The inset of Fig.~\ref{fig:mfig2} shows the order parameter as a function 
of $Q$ for all networks studied here. 
These results suggest that $Q_1$ depends basically on the dimension of the 
initial lattice. Namely, $Q_1=0.068(6)$ for all networks 
derived from a two-dimensional lattice independently of the value of 
coordination number. On the contrary, $Q_1$ is much smaller 
for the graph derivated from the cubic lattice and substantially 
larger for the network originated from the ladder-shape (practically 
one-dimensional) lattice.

The second phase transition (from the state $O$ to $A$) can be studied
more efficiently (or with a higher accuracy) if varying $P$ rather than
$Q$ because the limit cycle is affected by the quenched structural
randomness even for large $N$. This discrepancy can be avoided by
averaging over many runs on different structures whose creation is very
time-consuming. For temporary randomness ($P>0$), however, the technical
difficulties can be overcome more easily and the numerical analysis is 
executable with an adequate accuracy. This is the reason why henceforth
we concentrate on the second phase transition occurring with the increase
of $P$.

\begin{figure}
\centerline{\epsfig{file=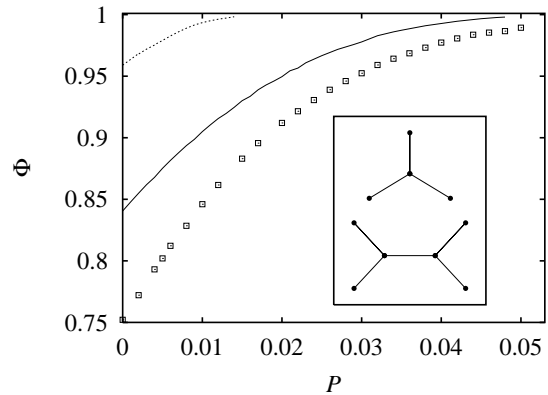,width=8cm}}
\caption{\label{fig:mfig3}The order parameter as a function of $P$ 
for the three-degree random regular graph. 
Boxes indicate the results of MC simulation. The estimated critical value 
of the transition to absorbing state is $P_2=0.079(2)$.
The dashed (solid) line represents the prediction of dynamical cluster
technique at the levels of four- and six-site approximation. Inset shows
the shape of four- and six-site clusters used in these approximations.}
\end{figure}

First we study what happens on random regular graphs (limit $Q \to 1$)
for $z=3$ when increasing $P$. This choice was motivated by the simplicity
of this tree-like structure on which the dynamical cluster technique can
be applied for sufficiently large clusters. In Figure \ref{fig:mfig3} the
MC results show that the order parameter $\Phi$ tends very slowly to 1.
Despite the large sizes in simulations we couldn't study the very close
vicinity of the transition point ($P_2$) because of the fluctuations
yielding an occasional transition from the noisy limit cycle to one of
the homogeneous states ($A$). Our MC data can be well approximated by a
power law behavior, namely $1-\Phi \propto (P_2-P)^\gamma$ with 
an exponent $\gamma=3.3(4)$. 

\begin{figure}
\centerline{\epsfig{file=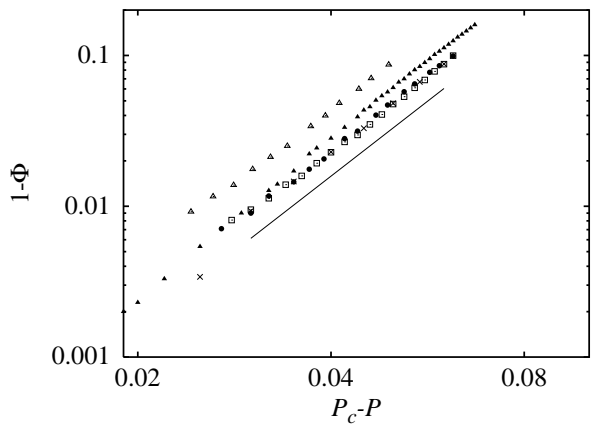,width=8cm}}
\caption{\label{fig:mfig4}Comparison of the continuous transitions to 
absorbing phase for different networks.
The MC data are obtained on honeycomb (triangle), square (cross),  
triangle (bullet) lattices, and on random regular graph with a degree of
$z=3$ (box). Filled triangle represents the result of dynamical cluster
approximation at six-site level for the same graph.
The solid line represents the slope of $\beta=3.3$.}
\end{figure}

For periodic structures the dynamical cluster technique is proved to be
a very efficient method to describe different phenomena in several
non-equilibrium models
\cite{dickman:pra90,szabo:pra91,szabo:jpa04,marro:99}.
In the limit $N \to \infty$ the random regular graph become locally
tree-like and it can considered as a Bethe lattice on which this technique 
works well too \cite{szabo:pre00b}.
Using this technique one can determine the probability of each configuration
occurring on a $k$-site cluster by solving a suitable set of master
equations (details are given in \cite{dickman:pre01,szabo:jpa04}).
The cluster size $k$ is a crucial parameter. For one-site clusters this
technique is equivalent to the mean-field approximation predicting
concentric orbits independent of $P$ \cite{hofbauer:98}. The pair approximation
($k=2$) gives spiral trajectories reaching the edge of triangles (or
the absorbing states) for arbitrary $P$. 
Choosing four- and six-site clusters (see the inset in Fig.~\ref{fig:mfig3})
this method was capable to reproduce the appearance of limit cycles below
a threshold value of $P$. The quantitative predictions of this method are
compared with the MC results in Fig.~\ref{fig:mfig3}. Obviously, the
increase of cluster size improves the estimation. Notice that according to
the six-site approximation the order parameter $\Phi$ also tends to 1 very
slowly (see, Fig.~\ref{fig:mfig4}). The extrapolated critical values 
are $P_2^{(4s)} = 0.019(2)$ and $P_2^{(6s)} = 0.067(4)$.  

The above value of the exponent $\gamma$ agrees (within statistical
error) with those we have observed on the square lattice in our previous
work concentrated only on the effect of  annealed randomness.
This coincidence inspired us to study the robustness of this transition. 
For this purpose the MC analyses of the second transition were carried
out on some two-dimensional lattices (square, triangle, and honeycomb
for $Q=0$). The results, summarized in Fig.~\ref{fig:mfig4}, seem to confirm
that the transition from the global oscillation (phase $O$)
to the absorbing states ($A$) is universal.

To summarize, we have studied the effect of host lattice randomness
on the stationary state for a simple rock-scissors-paper system.
The quenched and annealed randomness of the regular background is 
characterized by two parameters $Q$ and $P$ varying from 0
(corresponding to a lattice with a dimension of 1, or 2, or 3)
to 1 (random regular graph and/or mean-field condition). This system
displays two subsequent transitions if the randomness is increased.
A self-organizing pattern can be observed if these randomness parameters
remain within a region of $P-Q$ plane. When crossing the boundary of
this region a global oscillation (limit cycle) occurs via a
Hopf-bifurcation. The transition point (for $P=0$) depends strongly
on the dimension $d$ of the original lattice meanwhile it is hardly
affected by the clustering coefficient.
For the global oscillation the area of limit cycle (as well as the
``amplitude'') increases with the randomness up to its saturation
value. Thus, above a second threshold value [more precisely for
$P>P_2(Q)$] the system sooner or later terminates in one of the
homogeneous absorbing states. Our simulations indicate that this
second transition has also a universal feature in the slow tendency
towards the saturation value. It also turned out that the global
oscillation is stable ($\Phi < 1$) on the quenched random
regular structures if the number of neighbors is not larger than four 
($z \le 4$).
It would be interesting to see how the synchronization (as well as
the above mentioned two transitions) emerges on other random networks.
The present rock-scissors-paper model involves two crucial features.
On the one hand, the dynamical rule is cyclically symmetric; on
the other hand, the invasion fronts become very irregular even
on the two-dimensional lattices because the invasion between two
neighboring sites is not affected by their neighborhood. Further
investigations are required to clarify what happens if the dynamical
rules are not cyclically symmetric and/or the moving interfaces
are smoothed by local interactions. 

\begin{acknowledgments}

This work was supported by the Hungarian
National Research Fund under Grant No. T-47003 and 
Bolyai Grant No. BO/0067/00.
\end{acknowledgments}


\end{document}